\documentclass{webofc}
\usepackage[varg]{txfonts}   
\usepackage{color}
\usepackage{multicol}
  \usepackage{soul}

\begin{document}
\title{Understanding the cosmic abundance of $^{22}$Na: lifetime measurements in $^{23}$Mg}
%
%

\author{\firstname{C.} \lastname{Fougères}\inst{1,2}\fnsep\thanks{\email{cfougeres@anl.gov}} \and
        \firstname{F.} \lastname{de Oliveira Santos}\inst{1}\fnsep\thanks{\email{oliveira@ganil.fr}} \and
        \firstname{N. A.} \lastname{Smirnova}\inst{3}\and
        \firstname{C.} \lastname{Michelagnoli}\inst{1, 5}\and
          \firstname{} \lastname{GANIL-E710 / AGATA collaborations}
}

\institute{
Grand Accélérateur National d’Ions Lourds (GANIL), CEA/DRF-CNRS/IN2P3, Caen, France
\and
           Physics Division, Argonne National Laboratory, Lemont, USA
         \and
LP2IB, Université de Bordeaux, CNRS/IN2P3, Gradignan, France
         \and
Institut Laue-Langevin, Grenoble, France
        }

\abstract{%
Simulations of explosive nucleosynthesis in novae predict the production of $^{22}$Na, a key astronomical observable to constrain nova models. Its gamma-ray line at 1.275~MeV has not yet been observed by the gamma-ray space telescopes. 
The $^{20}$Ne/$^{22}$Ne ratio in presolar grains, a possible tool to identify nova grains, also depends on $^{22}$Na produced. Uncertainties on its yield in classical novae currently originate from the rate of the $^{22}$Na(p, $\gamma$)$^{23}$Mg reaction. At peak novae temperatures, this reaction is dominated by a resonance at E$_{\text{R}}$=0.204 MeV, corresponding to the $E_x$=7.785~MeV excited state in $^{23}$Mg. The resonance strengths measured so far disagree by one order of magnitude. An experiment has been performed at GANIL to measure the lifetime and the proton branching ratio of this key state, with a femtosecond resolution for the former. The reactions populating states in $^{23}$Mg have been studied with a high resolution detection set-up, i.e. the particle VAMOS, SPIDER and gamma tracking AGATA spectrometers, allowing the measurements of lifetimes and proton branchings. We present here a comparison between experimental results and shell-model calculations, that allowed us to assign the spin and parity of the key state. Rather small values obtained for reduced $M1$ matrix elements, $|M(M1)|\lesssim 0.5$~$\mu_N$, and proton spectroscopic factors, $C^{2}S_{\text{p}}$<10$^{-2}$, seem to be beyond the accuracy of the shell model. With the reevaluated $^{22}$Na(p, $\gamma$)$^{23}$Mg rate, the $^{22}$Na detectability limit and its observation frequency from novae are found promising for the future space telescopes.}
\maketitle
%
\par Many nuclei are synthesized in explosive stellar environments, like classical novae, supernovae and neutron star mergers. Novae are transient astronomical events and the most frequent explosive events in our galaxy, after X-ray bursts.
These explosions happen in a close binary stellar system consisting of a white dwarf accreting hydrogen-rich matter from its companion. This matter is progressively compressed at the white dwarf surface until the conditions required for hydrogen combustion are reached. Nucleosynthesis takes place during this explosive stage while a part of the new material is being ejected in the interstellar medium. Beyond this well understood picture, large uncertainties remain in our knowledge of novae, for example \color{black}
the white dwarf  initial conditions (mass, luminosity), the accretion rate and its composition, the amount of admixed white dwarf material with the accreta, the ejected mass, to list a few \cite{bookJose}. After the explosion, gas is crystallized, imprinting the ejecta composition into grains. 
The isotopic composition of presolar grains of a putative nova origin shed some lights into these stellar explosions \cite{bookJose, meteo, meteoAl}. Furthermore, novae are thought to end as supernovae of type Ia \cite{SNIaref} which are used as standard candles to determine distances and, so, to estimate the cosmic expansion. 
Therefore, astronomical observables are required to improve our understanding of novae: $^{22}$Na has revealed itself as a key candidate.

\par
\begin{figure}[h!]
\centering
\includegraphics[width=0.58\textwidth]{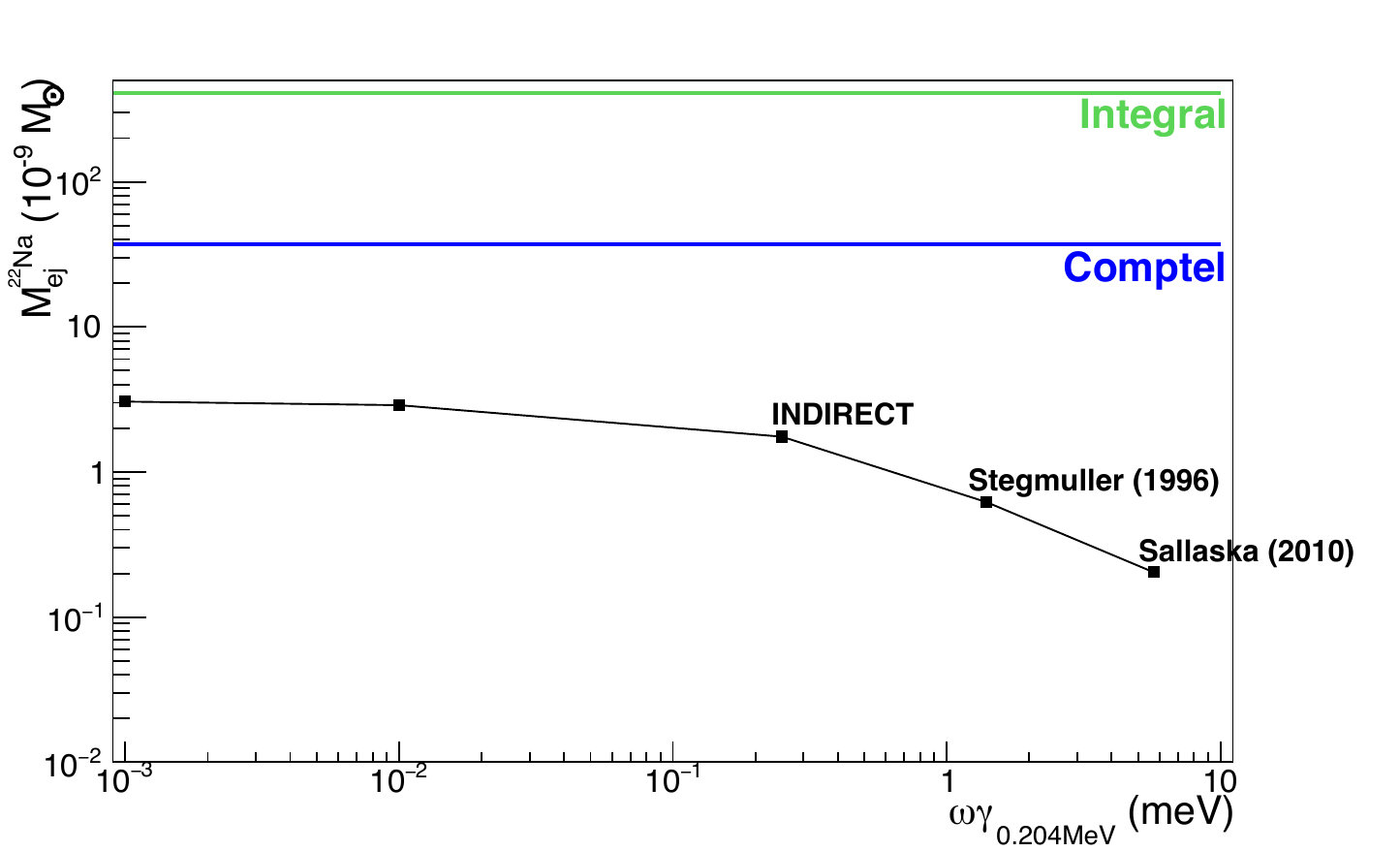}\includegraphics[width=0.43\textwidth]{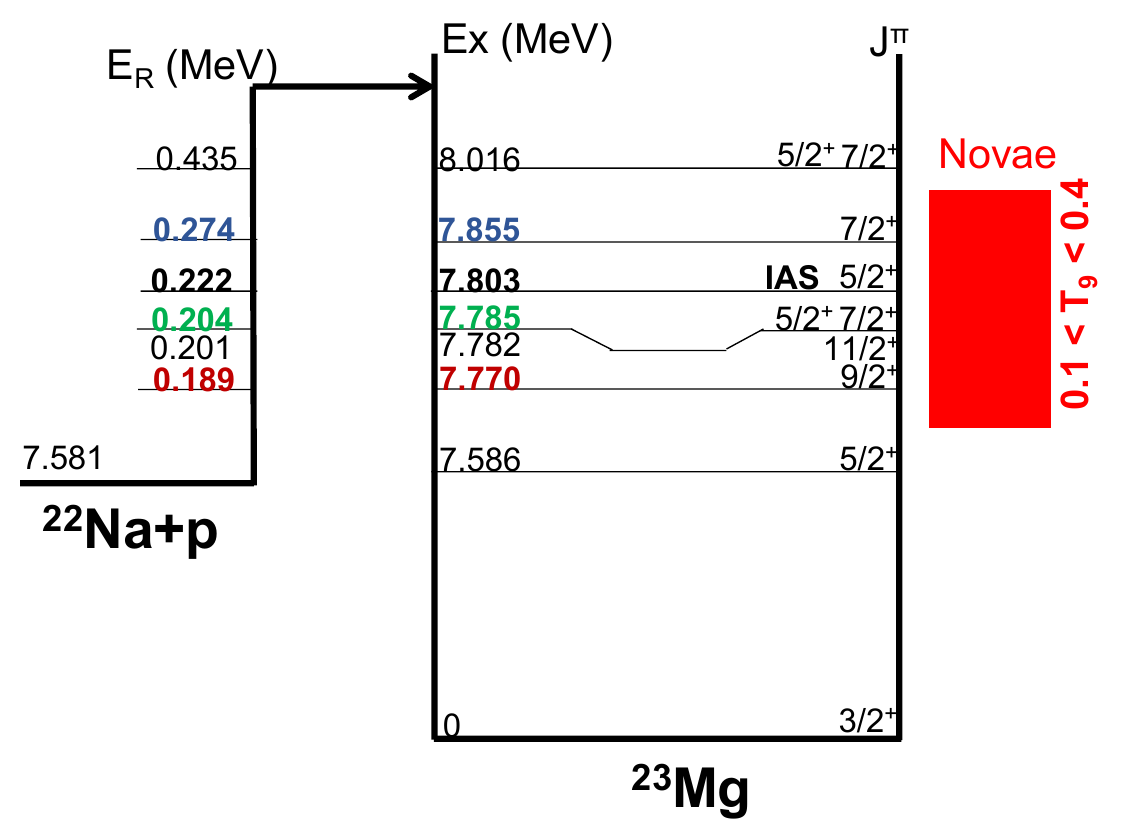}
\caption{{\bf{Left:}} The expected $^{22}$Na mass ejected from a nova is given   as a function of \color{black} the strength of the E$_{\text{R}}$=0.204~MeV resonance, from simulations with the {\tt MESA} code \cite{Paxton2013}. Colored lines mark the Integral and Comptel space telescope lower sensitivity limits. {\bf{Right:}} Levels scheme of $^{23}$Mg and resonance energies E$_{\text{R}}$ of the proton capture reaction. The Gamow window of novae is shown with the red zone.}
\label{figASTRO}
\end{figure}

 The lifetime of the $^{22}$Na radioisotope ($\tau$=2.6~yr) makes it a good radioactive tracer of novae~\cite{dhiel} since:
{\itshape{(i)}} the lifetime is longer than the duration of the opaque phase following the explosion (on timescales of hours), {\itshape{(ii)}} the decay time is short enough to ensure space-time correlation. Hence, $^{22}$Na is a promising candidate for observing $\gamma$-ray emissions from novae \cite{refNalifetime}. The emitted $\gamma$ rays at $E_\gamma$=1.275~MeV (and 0.511~MeV) are expected from hours to months after the explosion \cite{MHernanz}. They were searched for by the missions CGRO/COMPTEL~\cite{refCOMPTEL} and INTEGRAL/SPI~\cite{refINTEGRAL}, leading only to upper limits for the ejected amount of $^{22}$Na (left Fig. \ref{figASTRO}).
New space telescopes are in preparation, namely COSI~\cite{RefNextCSIastroGam} and e-ASTROGAM~\cite{refeASTRO}, with sensitivities 20 times higher than INTEGRAL. The radioisotope $^{22}$Na can be also measured, indirectly, by the over-abundance of the daughter nucleus $^{22}$Ne in presolar grains. Neon noble gas is not expected to easily condense into dust grains and, so, the presence of $^{22}$Ne implies in-situ $\beta^+$ decays of $^{22}$Na. A $^{22}$Ne excess was observed \cite{Black1972}. Therefore, it is critical to well quantify $^{22}$Na nucleosynthesis in novae.

\par 
ONe novae are predicted to be the main site of $^{22}$Na production \cite{Jose1999}. In these sites, the matter is composed of a mixture between the transferred proton rich gas and the white dwarf   envelope \color{black} with mainly oxygen and neon. The nucleosynthesis involves H to Ca isotopes, proton capture reactions and $\beta^+$ decays \cite{bookJose}. The production pathway of $^{22}$Na is $^{20}$Ne(p,$\gamma$)$^{21}$Na(p,$\gamma$)$^{22}$Mg($\beta^{+}$)$^{22}$Na at high temperatures and $^{20}$Ne(p,$\gamma$)$^{21}$Na($\beta^{+}$)$^{21}$Ne(p,$\gamma$)$^{22}$Na otherwise. The thermal window of novae is [0.05, 0.5]~GK. The destruction of $^{22}$Na proceeds via $\beta^+$ decays and $^{22}$Na(p,$\gamma$)$^{23}$Mg. The latter reaction is the least known, resulting   in \color{black} a factor of 10 uncertainty in the predicted amount of $^{22}$Na ejected (left Fig. \ref{figASTRO}). Since the direct capture contribution was found to be negligible \cite{thesisFougeres, Stegmuller1996}, the total reaction rate is determined by several narrow resonances in the Gamow window 
$<\sigma v >_{\text{tot}} = \sum_i(\frac{2\pi}{\mu_{(^{22}\text{Na,p})}\text{k}_{\text{BT}}})^{\frac{3}{2}}\hbar^2(\omega\gamma)_i\exp(-\frac{\text{E}_{\text{R}i}}{\text{k}_{\text{BT}}})$ (right Fig.~\ref{figASTRO}). 
A direct measurement at astrophysical energies showed that a resonance at E$_{\text{R}} = 0.204$ MeV dominates the $^{22}$Na(p,$\gamma$)$^{23}$Mg rate \cite{Direct}. However, its contribution disagrees by a factor 3 or even more with other measurements~\cite{Stegmuller1996,Seuthe1990}. Such a direct measurement is complex due to the radioactive nature of the target. The resonance strength can be determined indirectly from the spectroscopic properties of the $E_x$=7.785~MeV state in $^{23}$Mg, i.e. $\omega\gamma = \frac{2\text{J}_{^{23}\text{Mg}}+1}{(2\text{J}_{^{22}\text{Na}}+1)(2\text{J}_{\text{p}}+1)} \frac{\Gamma_{\text{p}} \times \Gamma_{\gamma}}{\Gamma_{\text{tot}}}$. The total and proton widths can be experimentally accessed via the state lifetime $\tau=\hbar/\Gamma_{\text{tot}}$ and proton branching ratio $BR_{p}= \Gamma_{\text{p}}/\Gamma_{\text{tot}}$. But the measured lifetime, $\tau$=10(3)~fs \cite{Jenkins2004}, is in contradiction with shell-model predictions, as   explained \color{black} hereafter. 

\par  
An experiment has been performed at GANIL facility, France, to strongly reduce the rate uncertainties of  $^{22}$Na(p,$\gamma$)$^{23}$Mg through the measurements of $\tau$ and BR$_{\text{p}}$. It is briefly presented here. Deduced spectroscopic properties are then discussed in relation with results from shell-model calculations. 

\subsection*{Experimental approach}
\vspace{-0.1cm}
The state of interest has been populated with a $^{24}$Mg beam of E=4.6~MeV/u onto a $^3$He-implanted gold target. The reactions $^3$He($^{24}$Mg, $^4$He)$^{23}$Mg$^*$ have been selected and measured with: the spectrometer VAMOS \cite{VAMOS1} for $^{4}$He identification and momenta measurement, the $\gamma$-ray tracking spectrometer AGATA \cite{AGATA2}, and the silicon telescope SPIDER for proton momenta measurements. Momenta reconstruction has been improved thanks to a small gas chamber placed in front of VAMOS. The $\gamma$-ray transitions and lifetimes have been measured for about 20 identified states in $^{23}$Mg$^*$. The branching ratios have been measured for the measured protons from states close to the proton emission threshold. Details can be found in Ref. \cite{thesisFougeres} and results will be
more extensively discussed in forthcoming publications.

\subsection*{Nuclear structure considerations}
\vspace{-0.1cm}
 
Insights from theory are now discussed in relation with the performed experimental study of $^{23}$Mg. 
The shell-model calculations are first briefly explained. Then, the measured and theoretical reduced M1 matrix elements of the $\gamma$-ray transitions are compared for several levels in $^{23}$Mg. We finally focus on the spectroscopic properties of the $E_x$=7.785 and 7.803~MeV states, the latter being the Isobaric Analog State (IAS) of the $^{23}$Al ground state ($J^\pi$=$\frac{5}{2}^+$). Among the calculated levels in $^{23}$Mg, the $E_x$=7.785~MeV state has not yet been clearly identified because of its uncertain spin value \cite{Jenkins2004, Kwag2021}. While identifying the best shell-model candidate for this state, its spin assignment and possible isospin-mixing with the IAS are discussed. 
\vspace{-0.1cm}
\subsubsection*{Shell-model calculations}
\vspace{-0.2cm}
\color{black}
We present here a theoretical analysis of the partial $^{23}$Mg level scheme, based on the $sd$-shell calculations with charge-dependent versions of USDA and USDB \cite{USD, USDCAB} (USDAcpn and USDBcpn) using the {\tt NushellX} code \cite{NUSHELLX}. In general, those interactions are known to provide a very good description of $sd$-shell nuclei. Assessment of shell-model states with observations is more challenging at higher energies  because of the increasing level density. This has still been achieved for the $E_x$$\sim$7.5-8~MeV states in $^{23}$Mg (details in \cite{thesisFougeres}). Excited states in $^{23}$Mg, including resonant ones, have thus been identified by comparing experimental energies, gamma decay branching ratios and $^{23}$Al $\beta^+$ feeding with theoretical quantities. 
Let us recall that a partial electromagnetic decay width $\Gamma_{\gamma}$ is related to the reduced $\gamma$-ray transition strength $B_\gamma$ via $\Gamma_{\gamma,(\text{L, i} \rightarrow \text{f})} = \frac{8\pi(\text{L}+1)}{\text{L}[(2\text{L}+1)\text{!!}]^2} (\frac{\text{E}_{\gamma,0}}{\hbar\text{c}})^{2\text{L}+1}B_{\gamma}$.   Then, the reduced M1 matrix elements $|M(M1)|$ can be derived with $|M(M1)| = \sqrt{(2J_i+1)B(M1)}$. \color{black} The proton decay width can be expressed as $\Gamma_{\text{p}}=\sum_{\ell,j} C^2S_p(\ell j)\times \Gamma_{ \text{s.p.}}(\ell j)$ where 
$C^2S_p (\ell j)$ refer to proton spectroscopic factors, $\Gamma_{\text{s.p.}}(\ell j)$ to 
the single-particle proton decay widths which are calculated from proton scattering cross-sections on a Woods-Saxon potential. 
\vspace{-0.1cm}
 
\subsubsection*{Reduced M1 matrix elements of the $^{23}$Mg states}
\vspace{-0.2cm}
\color{black}
\par
\begin{figure}[h!]
\centering
\includegraphics[width=0.6\textwidth]{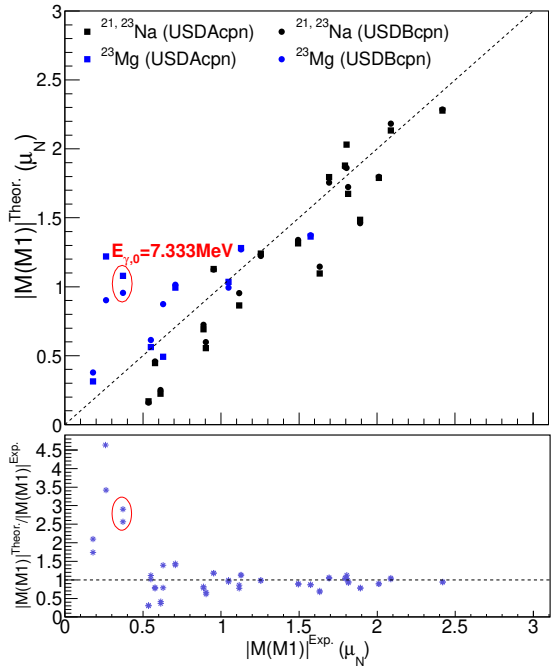}
\caption{{\bf{Correlation between measured and theoretical reduced matrix elements $|M(M1)|$ values for transitions in $^{23}$Mg and $^{21,23}$Na \cite{BM1NaIso}}}. Theoretical values were obtained from $sd$-shell calculations with USDAcpn and USDBcpn \cite{USD,USDCAB}, experimental ones from measured ($\tau$, E$_{\gamma,0}$), resulting in  rms=0.322~$\mu_N$ (USDAcpn) and rms=0.273~$\mu_N$ (USDBcpn). It follows that the largest relative deviations are observed for the lowest $|M(M1)|$ values, as 
for the $E_x$=7.785~MeV state shown in the red contours.}
\label{figBM1}
\end{figure}

 Experimental $B(M1)$ values have been determined from {\itshape{(i)}} measured ($\tau$, E$_{\gamma,0}$), {\itshape{(ii)}} identified M1 transitions \cite{nndc2022} among observed $\gamma$ rays and {\itshape{(iii)}} the assumption of small E2 contributions. The comparison of experimental and theoretical values of reduced M1 matrix elements is presented in Fig.~\ref{figBM1} for transitions in $^{23}$Mg and $^{21, 23}$Na \cite{BM1NaIso}. Optimum effective $g$-factors USDA(6) and USDB(6) from Table~I of Ref.~\cite{USDAB_em} have been used to parameterize the M1 operator.
The results for the state of interest are emphasized by the red contours. At the bottom, the ratios of shell-model to  experimental reduced matrix elements highlight differences by about a few times for small values of $|M(M1)|$. 
Although the accuracy in reproduction of $|M(M1)|$ for selected isotopes close to $^{23}$Mg is consistent with the rms deviations of about 0.22 - 0.20~$\mu_N$ reported for USDA and USDB~\cite{USDAB_em}, we see that the M1 transition from the key state of interest is among the most challenging for theory.
\vspace{-0.3cm}
 
\subsubsection*{Spectroscopy of the $E_x$=7.785 and 7.803~MeV states}
\vspace{-0.2cm}
\par  The two states measured at $E_x$=7.785 and 7.803~MeV (IAS) are now discussed in more   detail. \color{black} The $\gamma$-ray width, never measured for the IAS, is 3.0(2)~eV for the mirror state, in agreement with the shell-model results of 2~eV (USDAcpn) and 3~eV (USDBcpn).
From the previous experiment on BR$_{\text{p}}$ \cite{Perjarvi}, the proton width and spectroscopic factor are deduced for this isospin-forbidden proton emission, i.e. $\Gamma_{\text{p}}$=5(2)~meV and $C^{2}S_{p}$=0.0056(20) versus $C^{2}S_{p}$=0.0073 (USDAcpn) and 3$\times10^{-5}$ (USDBcpn).  
\begin{figure}[h!]
\centering
\includegraphics[width=0.89\textwidth]{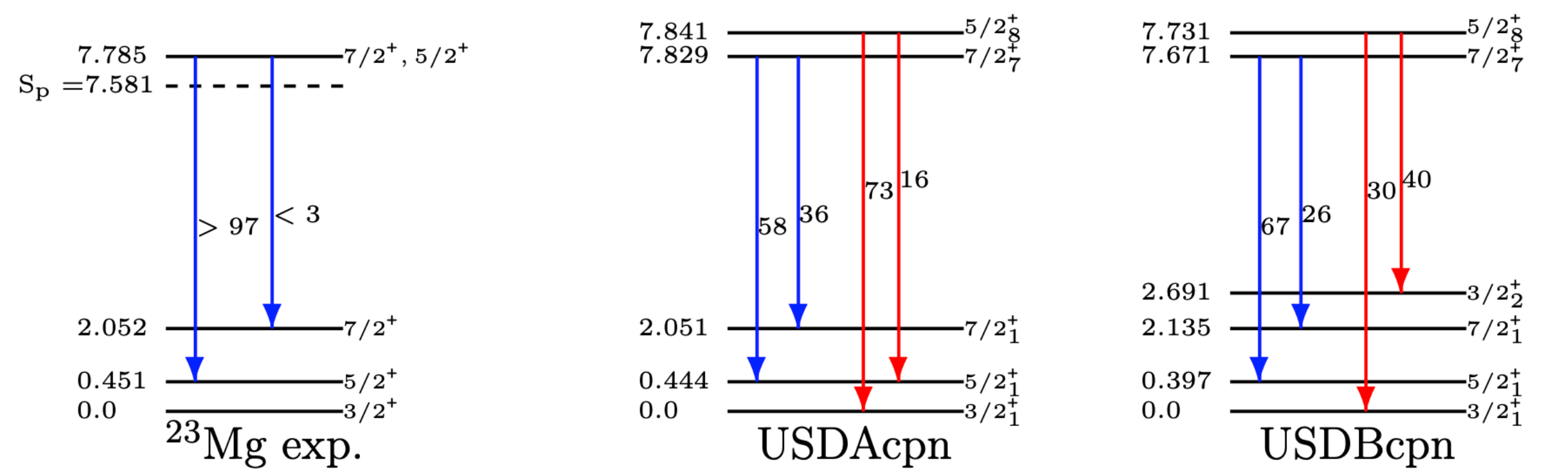}
\caption{{\bf{Comparison between $\gamma$-ray transitions from the $E_x$=7.785 state and from shell-model candidates: (n$_{\text{SM}}$=7, $J^\pi$=$\frac{7}{2}^+_7$) and (n$_{\text{SM}}$=8, $J^\pi$=$\frac{5}{2}^+_8$)}}. A good agreement is observed for the blue transitions, strengthening the $\text{n}_{\text{SM}}$=7 candidate as the $E_x$=7.785~MeV state (intensities are given in \%.)}
\label{DecaySchemeUnknownJ}
\end{figure}
 
\par Three different spins were proposed for the $E_x$=7.785~MeV state, either $\frac{7}{2}^+$\cite{Jenkins2004} or ($\frac{3}{2}^+$,~$\frac{5}{2}^+$)\cite{Kwag2021}. The value $\frac{3}{2}^+$\cite{Kwag2021} is unlikely since this state was well observed in the beta-delayed proton emission of $^{23}$Al ($\frac{5}{2}^+$)\cite{Saastamoinen2011, Friedman2020}. 
The spin of $^{22}$Na ground state being  $3^+$, the transferred angular momentum should be at least $\ell=2$, that is too high to be competitive with the $\gamma$ emission. As shown in Fig.~\ref{DecaySchemeUnknownJ} (in red), the theoretical $\gamma$-decay pattern of the $\frac{5}{2}^+_8$ state (eighth predicted state, $\text{n}_{\text{SM}}$=8) strongly differs from the observed one. On the contrary, the $J^\pi$=$\frac{7}{2}^+_7$ state (in blue) agrees well with the measured decay pattern. 
\color{black}
This is true regardless of the interaction used. Moreover, a spin assignment $\frac{5}{2}^+$ of the key state would imply a strong isospin-mixing with the IAS (as predicted by the shell model) which is just 18~keV above. However, the measured  log$_{10}$(ft)=3.305(23) of the IAS \cite{zhai2007} is in excellent agreement with no mixing hypothesis, and the one of the $E_x$=7.785~MeV state agrees with the calculated  $\frac{7}{2}^+_7$ state. Consequently, $J^\pi$=$\frac{7}{2}^+$ is 
assigned for the key state.

\par This conclusion of no mixing is supported by the small gamma width of the key state, $\Gamma_{\gamma}\lesssim 0.1$~eV, if compared to the gamma width of the IAS. Regarding proton emission, from the measured $BR_{p}$, a small spectroscopic factor $C^{2}S_{p} \approx10^{-3}$ was obtained. 
Theoretical predictions are $C^{2}S_{p}$=0.003 (USDAcpn) and 0.05 (USDBcpn), assuming $7/2^+$ assignment. Although the former result captures the order of magnitude, we remark that similar to M1 observable, very small values of spectroscopic factors may not be perfectly described.


 
\subsection*{Summary}
\vspace{-0.1cm}
\color{black}
\begin{figure}[h!]
\centering
\includegraphics[width=6.5cm]{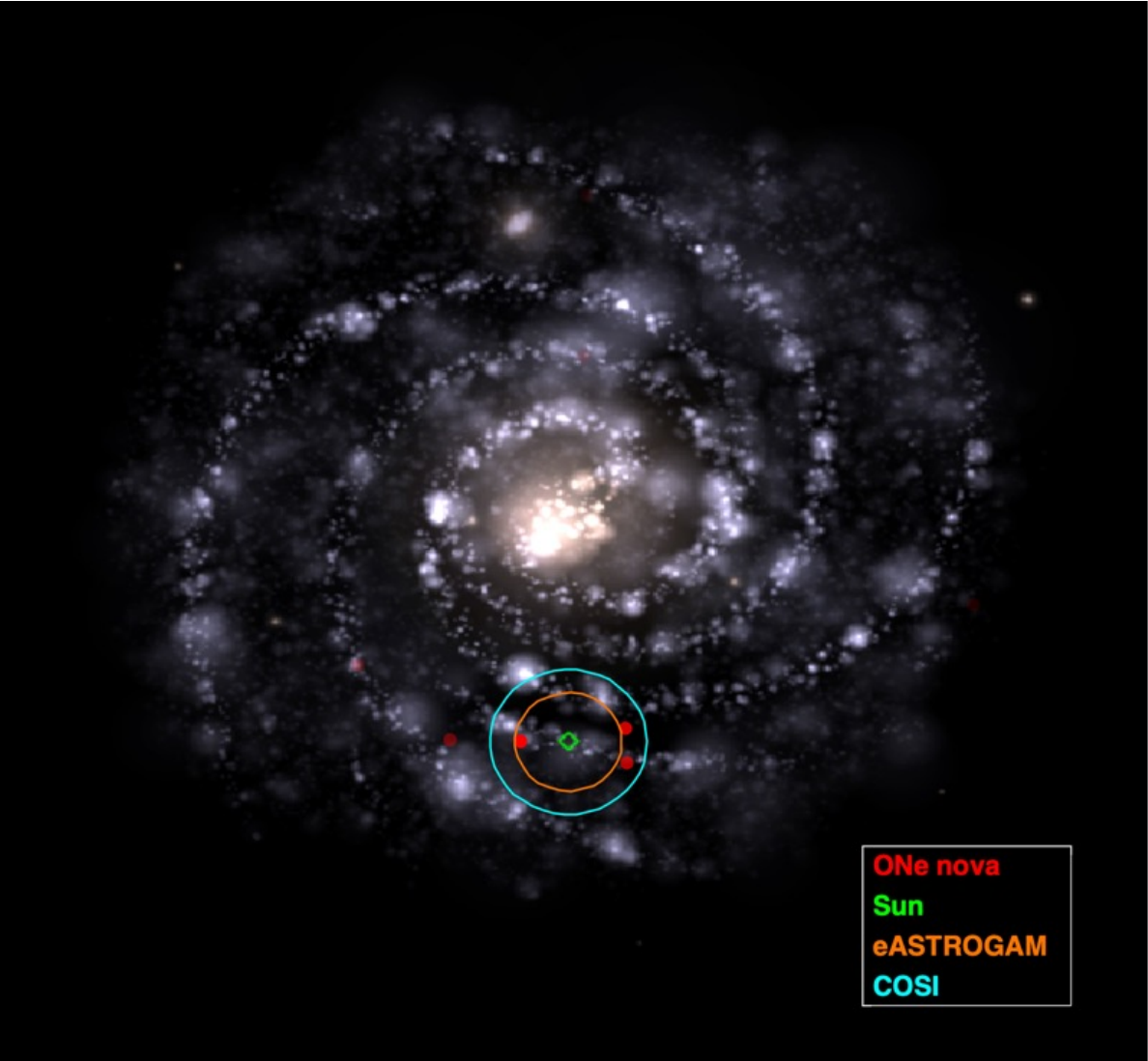}
\caption{{\bf{Map of the Milky Way with the expected flux of $^{22}$Na, represented by the red dots, from the ONe novae observed for the last 60 years.}} Circles represent the limit distance of COSI and eASTROGAM to detect $\gamma$-rays from $^{22}$Na.  The former should detect such emission once every 20~yr.}
\label{figNaMap}
\end{figure}
\par In the present work, a detailed comparison has been drawn between experimentally observed and shell-model states in $^{23}$Mg. It appears that shell-model predictions may have large relative uncertainties, up to a factor of 10, for very small (<10$^{-2}$) spectroscopic factors and M1 transition probabilities. This is apparently the case of the key state. Nevertheless, shell model has brought helpful arguments favoring $\frac{7}{2}^+$ spin assignment of the E$_{\text{R}}$=0.204~MeV resonant state. The measured total and partial widths of this state have led to a reevaluation of the $^{22}$Na(p,$\gamma$)$^{23}$Mg rate at novae temperatures. According to the future COSI and eASTROGAM sensitivities, it allows us to estimate the maximum detectability distance and a conservative observation frequency of $^{22}$Na based on the detected ONe novae \cite{surveyONe, bookJose}. These predictions, illustrated in Fig. \ref{figNaMap}, are encouraging the search for $^{22}$Na with the new space telescopes.
%
%
%

\subsection*{Acknowledgments} 
\vspace{-0.2cm}
The authors thank the GANIL accelerator staff for their beam delivery and their support, the AGATA collaboration, and the Normandie region.

\end{document}